# TOWARDS REDUCING FOREIGN LANGUAGE ANXIETY USING LEVEL-APPROPRIATE EMBODIED CONVERSATIONAL AGENTS

**K. Rajaratnam[1], W. Gan[2], Y. Sun[3]**

[1]*University of Oxford (UNITED KINGDOM)*
[2]*National Institute of Information and Communications Technology (JAPAN)*
[3]*National Institute of Informatics (JAPAN)*

## Abstract

Foreign language anxiety (FLA) can be a major barrier to second language acquisition (SLA), especially in conversational contexts. With the proliferation of large language models (LLMs) throughout all areas of life, recent work suggests that interacting with LLM agents can be instrumental within the field of SLA and foreign language education, especially for reducing FLA. Related work also suggests that linguistic demands and task complexity can be major predictors of FLA, implying that the use of demanding, complex language could lead to learners experiencing higher levels of FLA.

In this paper, we propose a novel multi-agent embodied conversational system that generates level-appropriate dialogue for English language learners. These levels are based on those defined by the Common European Framework of Reference for Languages (CEFR) to describe non-native listener and speaker proficiency. Using a “generate-evaluate-regenerate” loop with multiple LLM agents and a level classifier, it achieves a desired simplicity that is adaptive to the user’s proficiency level. We also share the results of a preliminary small-sample pilot study that tested this system with Japanese university students, to see whether it would yield lower FLA levels than an unsimplified embodied conversational agent.

Analysis of conversational output showed that 87.4% of dialogue sentences generated by the proposed multi-agent system fell within one predicted CEFR level of the learner’s self-assessed proficiency, compared to 54.1% for the unsimplified agent. This suggests that the novel system is better able to produce output at an appropriate level for the learner. Although the pilot study did not yield statistically significant evidence that the system reduces FLA levels in Japanese learners of English, likely due to its small sample size, it provided important usability findings and culturally-informed design insights that will inform the development of a larger-scale study.

## 1 INTRODUCTION

Originally observed within classroom environments in 1986, Horwitz et al. defined foreign language anxiety (FLA) as a specific anxiety associated with language learning environments [1]. This anxiety can make students apprehensive, stressed, or embarrassed when trying to communicate. This can sometimes lead students and learners to avoid speaking altogether, greatly hindering their improvement in expressing themselves in their target language. This anxiety is understood to be largely driven by three contributing factors: communication anxiety, test anxiety, and a fear of negative evaluation [1]. Horwitz et al. proposed the foreign language classroom anxiety scale (FLCAS) and an associated 33-item questionnaire to measure this phenomenon among language learners.

### 1.1 Reducing FLA with Conversational Agents

Conversational agents driven by large language models (LLMs) have been used as tools to assist in second language acquisition (SLA) and foreign language acquisition [2], and recent work even suggests that their use can be associated with lower FLA levels [3] [4] [5].

In 2024, Prakash proposed a Dutch tutoring system that aimed to address foreign-language anxiety with an “Episodic Memory” architecture powered by GPT-4 [4]. 28 Dutch learners were tasked with using the system and completing a questionnaire based on Horwitz et al.’s FLCAS scale. The resulting FLCAS scores suggested that this system has a strong potential to reduce speaking-related state anxiety, though participants shared various criticisms in a separate questionnaire, noting an “occasional mismatch between the content generated and their learning level.”

In 2025, Cheng et al. proposed an immersive English conversation system with an LLM-powered VR-embodied agent, with whom learners are tasked with approaching and initiating a short 3-minute “Icebreaker” conversation inside a virtual café [5]. Participants reported significantly lower FLA levels (measured using a modified version of Horwitz et al.’s FLCAS scale) after engaging in these "Icebreaker" sessions, when compared to the FLA they experience in the “real world.” Though these results seemed quite promising, the study suffered from a small sample size (*n* = 10) and a relative lack of diversity within the language-learning backgrounds of participants, all of whom lived in the English-speaking environment of London. Additionally, as in Prakash’s work, there is an unmitigated potential for inappropriate out-of-level content to be output by the agent.

### 1.2 Classifying Conversational Text by Target Proficiency Level

Previous work has also explored the classification problem of identifying the most appropriate Common European Framework of Reference (CEFR) level for a given text to assess its appropriateness for learners at different levels. In 2025, Kogan et al. released an expert-annotated dataset of short conversational English text samples distributed across the CEFR scale [6]. They found that a pretrained Bidirectional Encoder Representations from Transformers (BERT) model fine-tuned on their dataset outperformed other classification methods, including human expert classifiers. Despite their higher performance, a notable weakness of BERT-based classification is a lack of interpretability or explainability of predictions. Despite this “interpretability gap,” these classifiers could still be used as tools to better "level" the dialogue output of an LLM conversational agent for SLA purposes. Given this potential, a natural question arises as to whether speaking with an agent simplified in this manner has a tangible effect on learners' FLA levels. In this paper, we present a CEFR-level filtering approach to effectively create a “leveled” conversational agent and share preliminary findings from a small-sample pilot study conducted with Japanese learners of English.

## 2 METHODOLOGY

To create "leveled" conversational agents, Kogan et al.'s dataset was utilized to fine-tune a pretrained BERT model, while also providing a few-shot prompt to an LLM agent (gpt-oss with 20 billion parameters) tasked with giving feedback to guide simplification efforts for dialogue that has been determined to be too complex. This CEFR classifier was attached to an embodied LLM conversational agent (gpt-oss with 120 billion parameters) that is given a similar prompt and use-scenario to the Icebreaker agents in Cheng et al.'s work. The framework of this system is illustrated in Fig. 1.

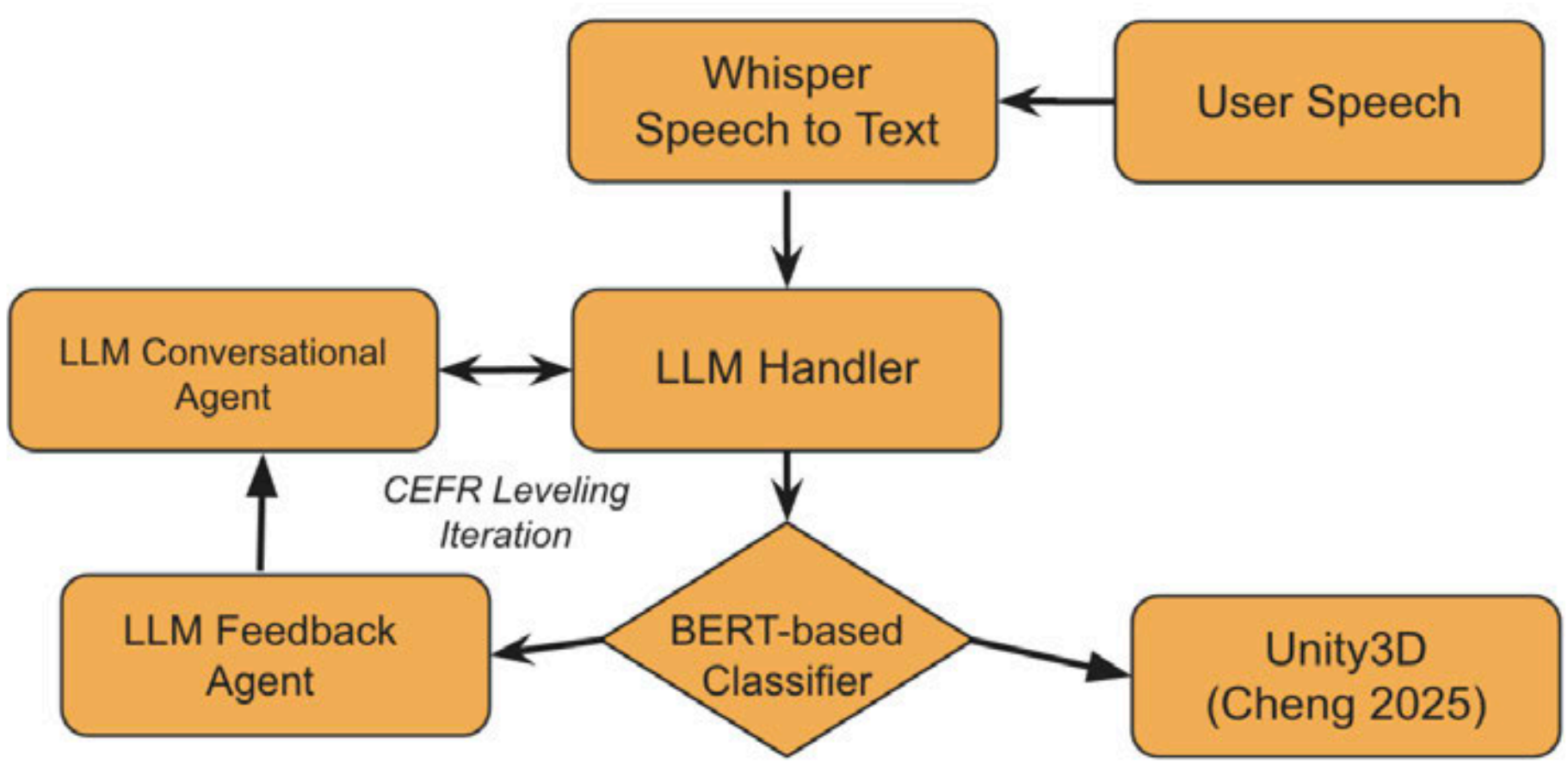

*Figure 1. Leveled Conversational Agent Framework*

### 2.1 Iterative CEFR Leveling Loop

Due to Kogan et al.’s dataset consisting of short conversational excerpts, the LLM’s initial output is first segmented into individual sentences, which are then classified into projected CEFR levels using a fine-tuned BERT model. This process can yield multiple CEFR predictions for a single response. While it might seem natural to compute an “average” CEFR level, doing so would be inappropriate given the ordinal nature of these categories.

Instead, we score each response by calculating the proportion of sentences at or below the target level, assigning half weight to sentences that are one level above the target to reward proximity. This score is then compared against a threshold ε (set to 0.66 in our experiments). If the score exceeds ε, the response is returned to the user and stored in the agent's memory.

If the score falls below ε, a separate, non-conversational LLM-based "feedback agent" generates an explanation by interpreting the CEFR predictions produced by the BERT classifier, helping to bridge the interpretability gap. It also suggests ways to simplify the response to better match the target level. This feedback is then provided to the original conversational agent, which generates a revised, simpler response while maintaining conversational coherence.

This process repeats until a response meets the threshold or a maximum number of iterations is reached. Notably, the agent's contextual memory retains only the final accepted responses, discarding intermediate feedback and revisions to preserve context window capacity for the ongoing conversation.

*Table 1. Conversational openers filtered to different CEFR levels by the leveled conversational agent*

| ***CEFR Filter Setting*** | ***Output Dialogue**** |
|---|---|
| A1 | "Hello. How are you? Do you like coffee?" |
| A2 | "Hey there! I'm enjoying this coffee. How's your day going?" |
| No Filter | "Hey there– just grabbing my morning mocha. How's it going?" |

** This is sample dialogue that was generated outside of any interactive sessions with participants.*

## 2.2 Pilot Study Design

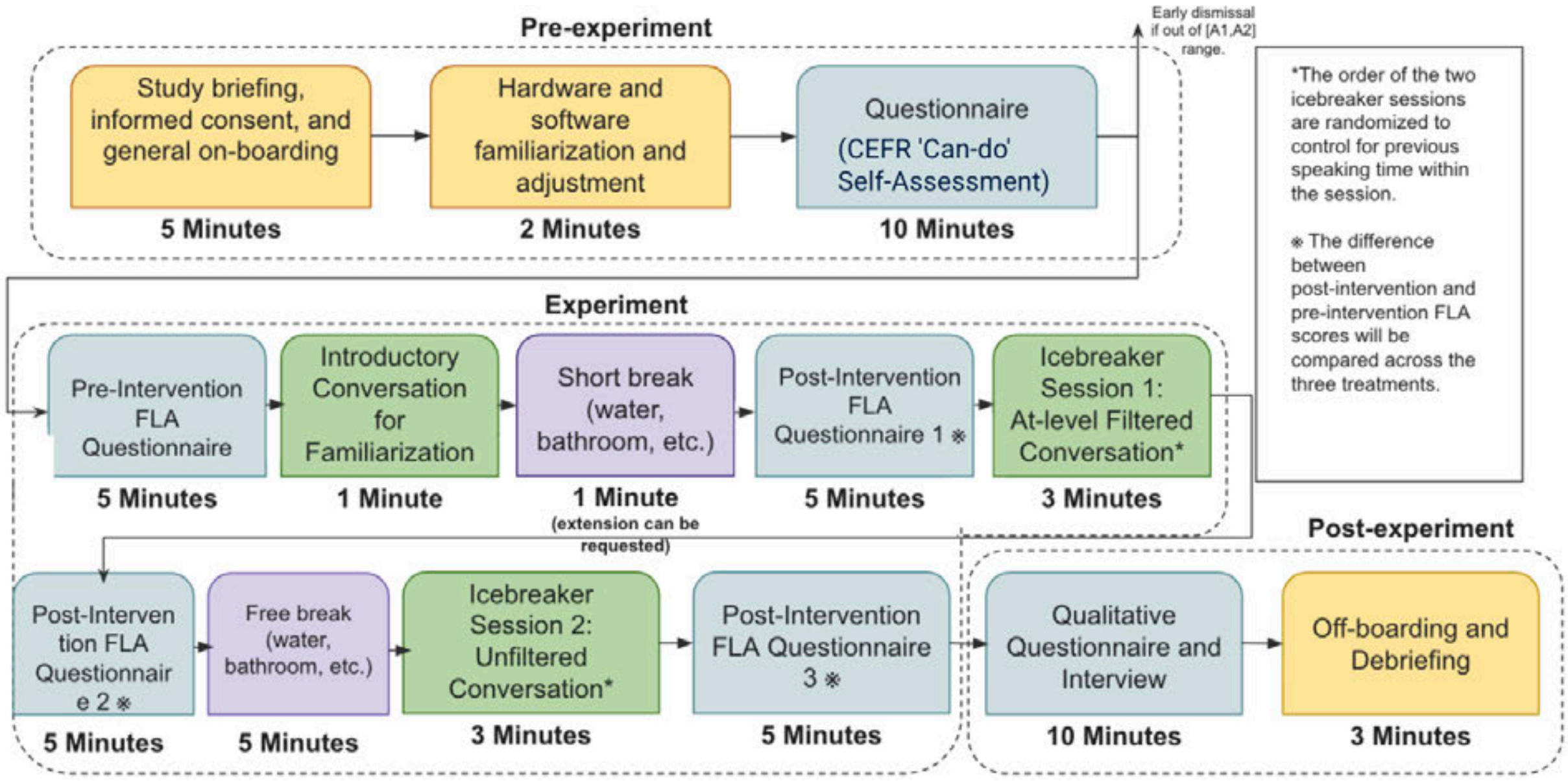

*Figure 2. Overview of Experimental Study Procedure*

The pilot study, whose overall procedure is depicted in Fig. 2, involved three Japanese first-year university students with prior experience learning English. All participants self-identified as "basic" users of the language, corresponding approximately to CEFR levels A1–A2. This proficiency range was selected because the proposed leveled filter is expected to have its strongest impact at lower levels, where linguistic complexity may more directly influence comprehension and anxiety.

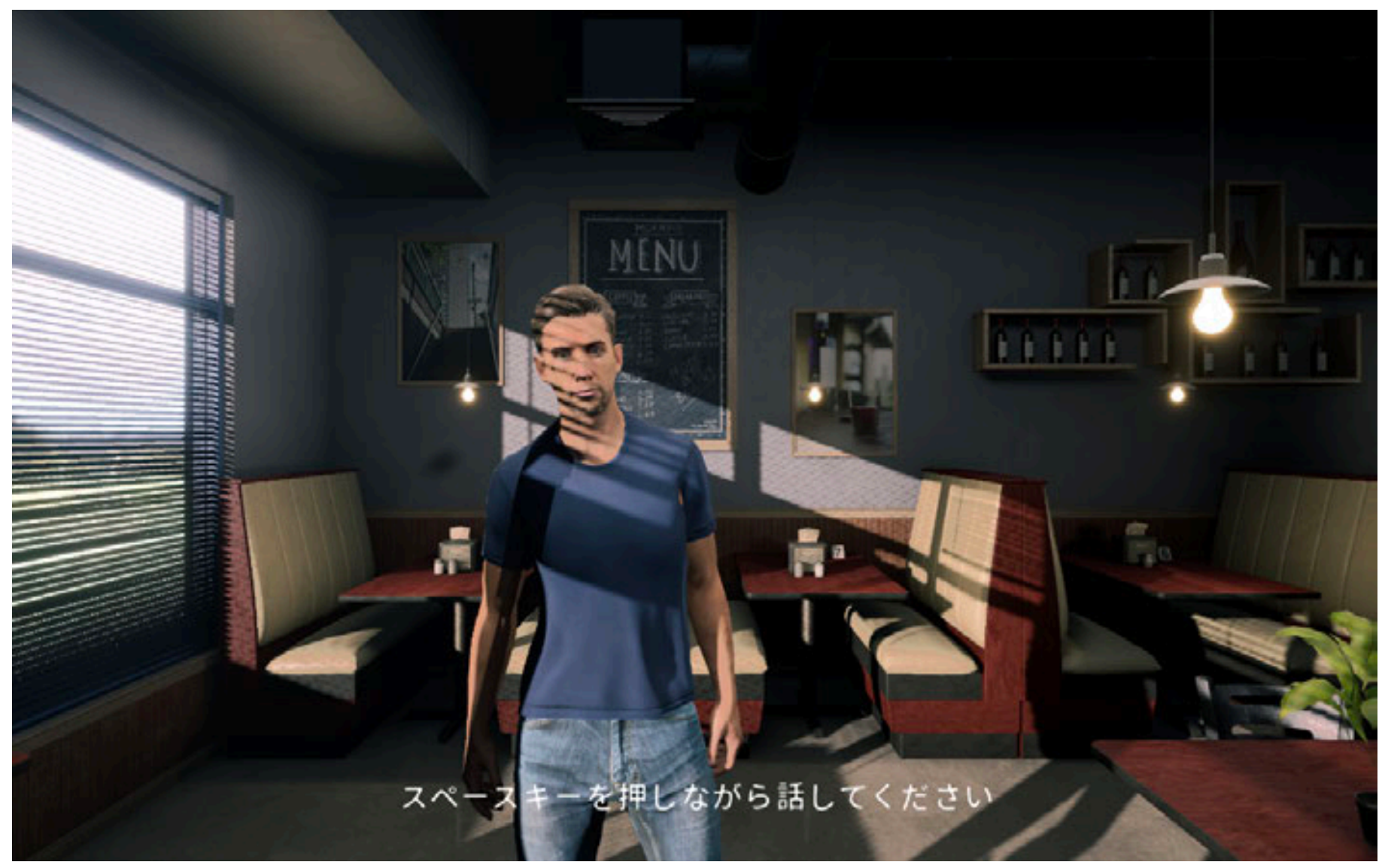

*Figure 3. Immersive User Speaking Interface with Embodied Conversational Agent*

Given the exploratory nature of this pilot study, a within-subject crossover design was used to examine preliminary effects of both filtered and unfiltered "Icebreaker" conversations. As illustrated in Fig. 3, interactions take place through a virtual agent embodied in a Unity3D scene resembling a café, identical to that used by Cheng et al. Each participant completes three conversational sessions. To reduce the potential impact of novelty-induced anxiety, the first session is always filtered and limited to a brief 1-minute familiarization task. The second and third sessions constitute the primary comparison conditions, with one filtered and one unfiltered conversation presented in randomized order, both 3 minutes long. While this design choice helps mitigate initial anxiety, it may introduce an order effect, which is acknowledged as a limitation. The timers only begin once the participant has successfully spoken into the microphone and begun the conversation. After the time has elapsed, the agent is prompted to naturally end the conversation with a transition response.

Following each conversation, participants complete a 15-item foreign language anxiety questionnaire adapted from the most relevant items of the Foreign Language Classroom Anxiety Scale translated into Japanese. Responses are recorded on a 4-point Likert scale, excluding a neutral midpoint to reduce central tendency bias commonly observed in Japanese respondents. Participants are then given a short break before proceeding to the next session.

Prior to the conversational tasks, participants' perceived CEFR levels are assessed using speaking and listening self-evaluation measures adapted from the Tokyo University of Foreign Studies [7]. Self-assessment was selected to capture learners' perceived communicative competence, which may interact with experienced anxiety, although it does not replace objective proficiency measurement. This self-assessment verified whether participants agreed to 79% or more of the can-do statements associated with each level, skipping items when appropriate. The self-assessment was divided into independent listening and speaking sessions; the lowest of the two levels was used to calibrate the leveling filter. Participants are also asked to complete a pre-intervention FLA questionnaire to serve as a baseline assessment.

Due to the small sample size, the study emphasizes feasibility and exploratory insights rather than generalizable outcomes. To complement the quantitative measures, qualitative data are collected through a free-response questionnaire and a semi-structured interview conducted after all sessions. These responses are analyzed to identify recurring themes related to user experience, perceived difficulty, and anxiety.

## 3 RESULTS

Table 2 presents participants' foreign language anxiety (FLA) scores before and after each treatment, along with their self-assessed listening and speaking abilities. Due to the small sample size ($n$ = 3), no statistically significant differences were observed. However, a tentative trend suggests that the filtered condition is associated with lower FLA for Participants 1 and 2. Participant 3 did not follow this pattern. FLA scores were calculated by adding all numerical responses for each item in the questionnaire, inverting the numbering of responses for reverse-coded items. Additionally, all participants reported the highest FLA before any conversations and immediately after the introductory session.

*Table 2. Foreign Language Anxiety Scores by Participant*

| | *Self-AssessedListening CEFR Level* | *Self-AssessedSpeaking CEFR Level* | *Pre-intervention FLA Score* | *FLA Score After Intro Conversation* | *FLA Score After Filtered Conversation* | *FLA Score After Unfiltered Conversation* |
|---|---|---|---|---|---|---|
| ***Particip.1*** | A1 | A1 | 42 | 39 | 36 | 38 |
| ***Particip.2*** | A2 | A1 | 43 | 41 | 33 | 37 |
| ***Particip.3*** | A1 | A1 | 42 | 41 | 37 | 36 |

** Participant 2 experienced a different permutation from the other two participants, with the unfiltered conversation happening before the filtered conversation.*

Fig. 4 shows a scatter plot of reported FLA against the percentage of sentences classified within one level of the target proficiency level or below. The plot suggests a negative relationship between the proportion of level-appropriate sentences and reported FLA. Additionally, Participant 3, who reported slightly higher FLA after the filtered conversation than after the unfiltered conversation, did not exhibit as large a difference in level-appropriateness between the two conversations as other participants.

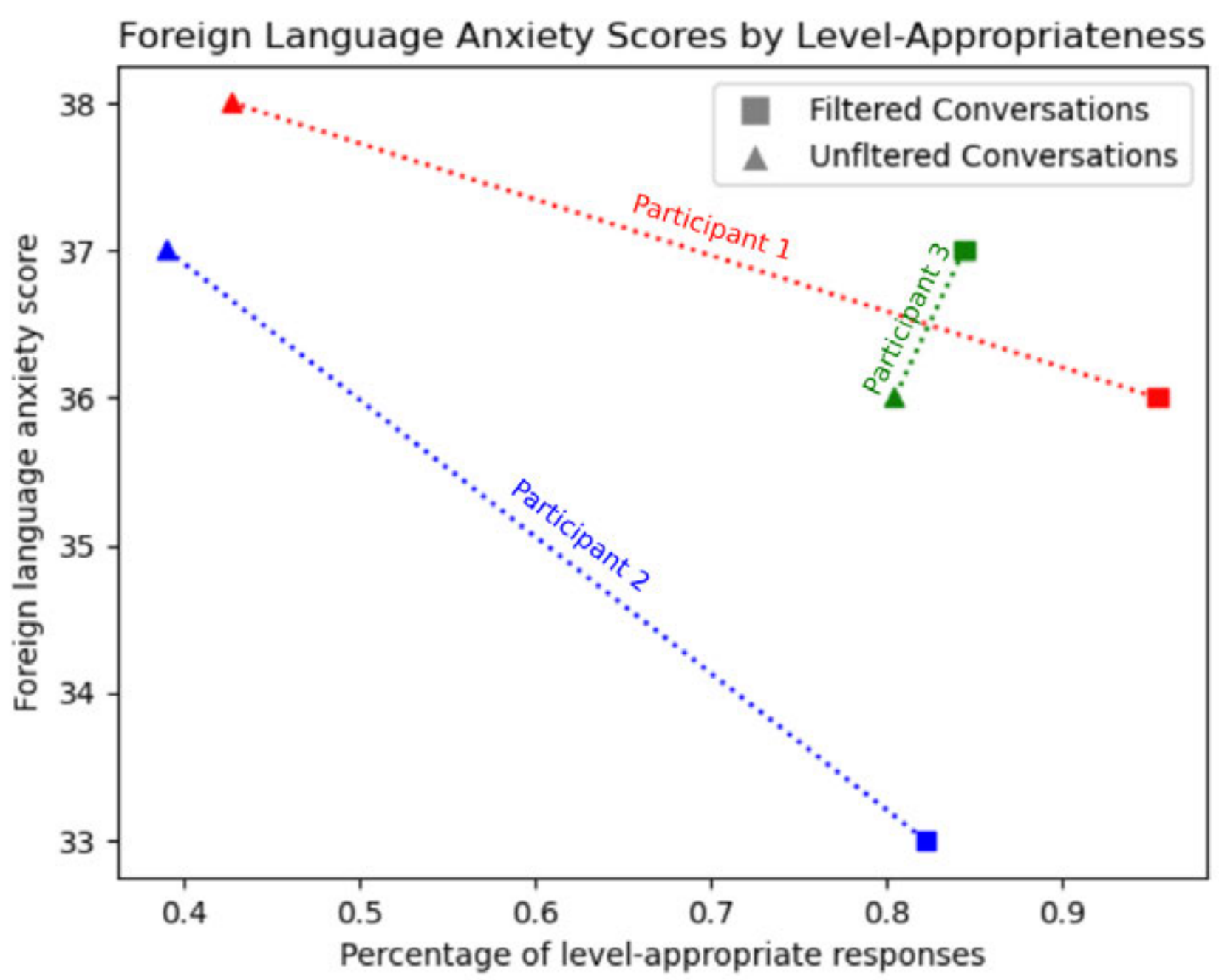

*Figure 4. Scatterplot of FLA scores versus Level-Appropriateness.*

Fig. 5 and Fig. 6 illustrate the distribution of CEFR classifications for sentences in the filtered and unfiltered conversations. Filtered conversations predominantly fell between the A1 and A2 levels and contained no sentences that were categorized above B2. Overall, 87.4% of sentences generated in the filtered condition fell within one predicted CEFR level of the learner's self-assessed proficiency, compared to 54.1% in the unfiltered condition.

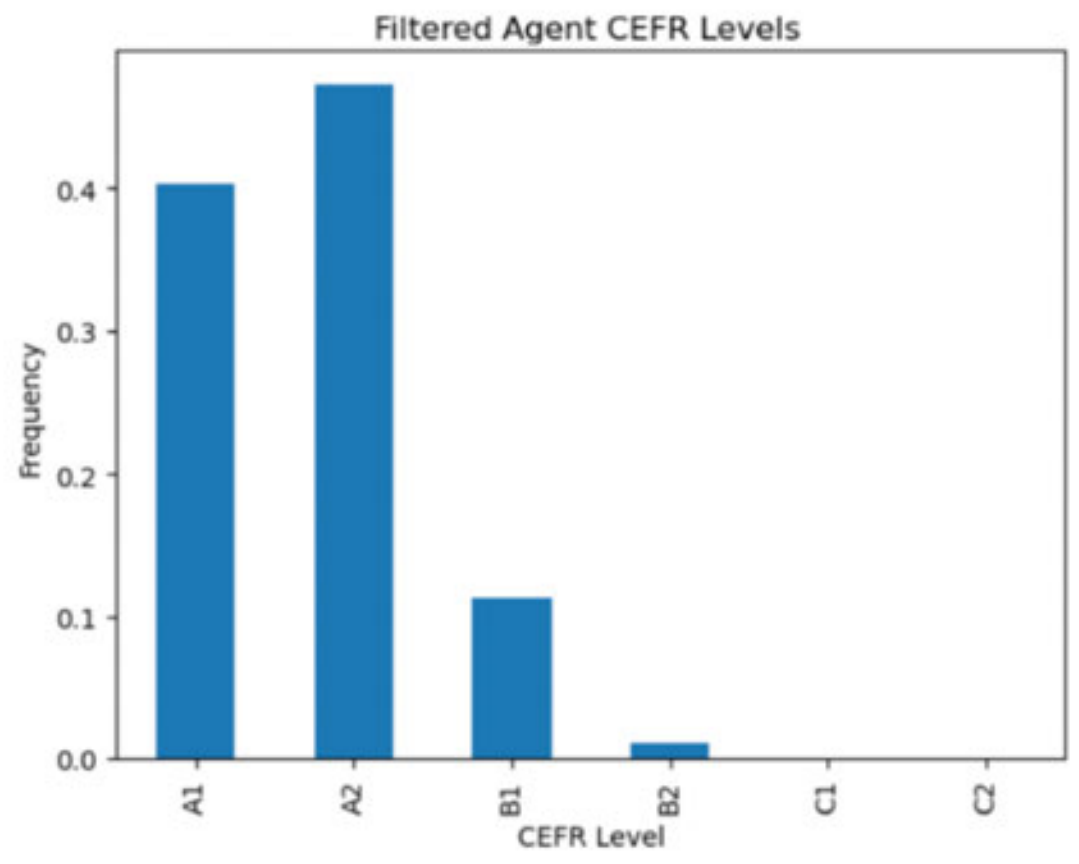

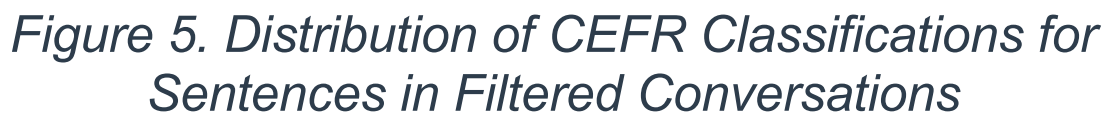
*Figure 5. Distribution of CEFR Classifications for Sentences in Filtered Conversations*

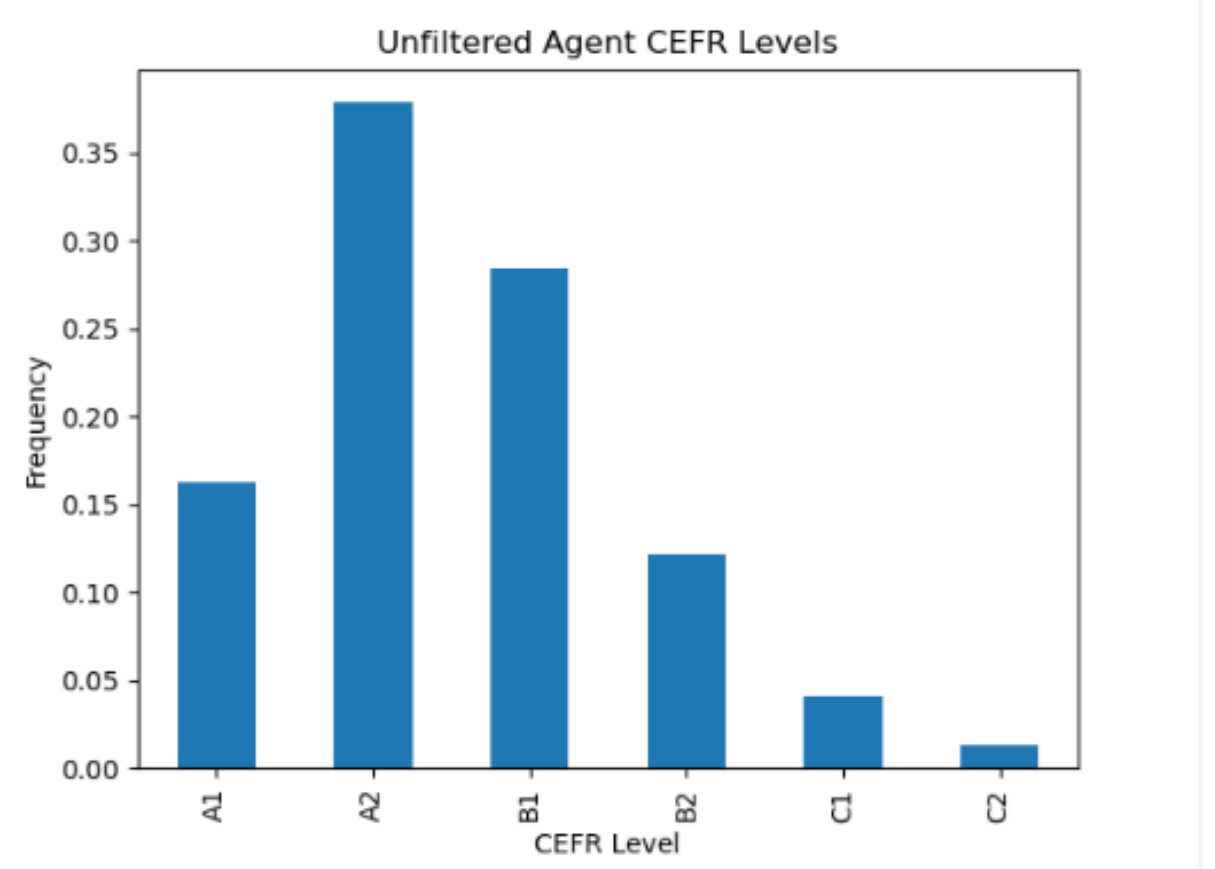

*Figure 6. Distribution of CEFR Classifications for Sentences in Unfiltered Conversations*

## 3.1 Qualitative Findings

Post-intervention interviews indicated that all participants found it easier to communicate with the conversational agent than with a native English speaker. However, several cultural and usability issues were reported.

Participants expressed uncertainty about the interaction scenario, particularly the task of approaching a stranger in a café. Some found this situation unnatural or unclear; one participant initially interpreted the role as that of an employee addressing a customer.

Only Participant 1 reported noticing a difference in language complexity between the filtered and unfiltered conditions. Despite this, all participants reported greater enjoyment in the filtered condition, including those who experienced it before the unfiltered condition.

Several usability challenges were also observed. Participants were sometimes not clearly detected by the microphone, and conversation topics varied across sessions. Some participants reintroduced themselves at the beginning of sessions, while others attempted to continue previous conversations. Participants were generally slow to initiate dialogue and appeared unfamiliar with the interface. In many cases, conversations were still developing when the three-minute time limit elapsed.

One participant described experiencing anxiety related to *ma* (間), referring to the pause between the agent's utterance and their response. Despite recognizing that the agent was not a real person, the participant reported feeling pressure to minimize these pauses.

# 4 CONCLUSIONS

This study investigated whether a filtered large language model–based conversational agent, designed to align its output with learners' proficiency levels, could help reduce foreign language anxiety (FLA). Although the small sample size ($n$ = 3) precludes any statistically significant conclusions, the findings provide preliminary insights into how linguistic alignment may influence learner experience.

Quantitative results showed that the filtered condition produced a substantially higher proportion of sentences within one CEFR level of participants' self-assessed proficiency compared to the unfiltered condition. While no statistically significant differences in FLA were observed, a tentative trend toward lower FLA in the filtered condition was evident for two of the three participants. The third participant did not exhibit this pattern; notably, during the unfiltered session, they reintroduced themselves at the beginning of the interaction, which appeared to constrain the conversation to a familiar and relatively simple topic. This suggests that conversational context, in addition to system-level filtering, may play a meaningful role in shaping perceived difficulty and anxiety. Supporting this interpretation, the observed negative relationship between the proportion of level-appropriate sentences and reported FLA indicates that both content complexity and topic familiarity may influence learner comfort.

Qualitative findings further contextualize these results. All participants reported that interacting with the conversational agent was less anxiety-inducing than speaking with a native English speaker, reinforcing the potential of LLM-based systems as low-pressure environments for language practice. Participants also consistently expressed a preference for the filtered condition, even when they did not explicitly perceive differences in linguistic complexity. This points to the possibility that reduced cognitive load or smoother interaction flow may contribute to improved user experience beyond conscious awareness.

At the same time, several limitations highlight important directions for future work. The interaction scenario of approaching a stranger in a café was perceived as culturally unfamiliar and, at times, confusing for participants based in Japan. Future systems could incorporate more culturally relevant and customizable scenarios to better match learners' backgrounds and expectations. Additionally, usability challenges such as microphone sensitivity, interface unfamiliarity, and strict time constraints appeared to limit participant engagement. Addressing these issues through improved interface design, adaptive conversation pacing, and longer or more flexible interaction periods may yield more representative results.

The role of conversational structure also warrants further investigation. The observation that a simple self-introduction influenced perceived difficulty suggests that future studies should more carefully control or systematically vary conversation topics. For example, comparing constrained (e.g., introductions) versus open-ended dialogues could help disentangle the effects of topic familiarity from those of linguistic complexity. Similarly, incorporating objective proficiency measures and logging interaction features such as response latency may provide a more nuanced understanding of how anxiety manifests during real-time interaction.

An important methodological consideration concerns the measurement of FLA. This study employed a Japanese translation of the Foreign Language Classroom Anxiety Scale (FLCAS) originally developed by Horwitz et al., which was adapted for use in this experimental context. However, prior research has noted that direct application of the FLCAS across cultural contexts may not fully capture the nuances of learner anxiety. For instance, Motoda proposed a revised scale tailored to Japanese learners in 2000, separating anxiety into in-class and out-of-class dimensions to better reflect differences in social expectations and communicative settings [8]. Given that the present study situates interaction in a quasi-real-world conversational scenario rather than a classroom environment, future work may benefit from adopting or extending such context-sensitive instruments to more accurately capture variations in FLA within a Japanese cultural context.

Finally, the emergence of the perceived pressure associated with response timing highlights the importance of temporal dynamics in human-AI interaction. Future work could explore prompting the agent to generate conversational hooks or utilize specific sentence patterns to help simplify the process of thinking about how to respond and to reduce pressure during conversational gaps.

In summary, this study offers preliminary evidence that proficiency-aligned conversational agents may help reduce FLA and improve learner experience, while also underscoring the importance of contextual, cultural, and interactional factors. Further research with larger and more diverse samples, improved experimental controls, and more culturally responsive system designs and questionnaires will be essential to more rigorously evaluate and extend these findings.

## ACKNOWLEDGEMENTS

This research was fully conducted on-site at the National Institute of Informatics (NII) office in Chiyoda, Tokyo, with all data storage, collection, and analysis occurring at the NII. This study fully complies with the NII's institutional ethical requirements (Approval No. R7-16-1).

This research was conducted over six months as a part of the NII's 2025-2026 International Internship Program, based on a memorandum of understanding with the University of Oxford.

We would also like to thank Zimu Cheng for providing us with the 3D art assets needed to reproduce the simulated environmental conditions of his study.